\documentclass{article}
\usepackage{graphicx} 
\usepackage[T1]{fontenc} 
\usepackage[style=authoryear,minbibnames=3,natbib=true]{biblatex}
\newcommand{\textapprox}{\raisebox{0.5ex}{\texttildelow}}

\title{Proof of Response}
\author{Illia Polosukhin, Alex Skidanov \\ NEAR AI \\ illia@near.ai, alex@near.ai}
\date{}

\begin{document}

\maketitle

\begin{abstract}
    We present a mechanism that for a network of participants allows one participant of the network (Alice) to request some data from another participant (Bob) and either receive a response from Bob within a known-in-advance, bounded time \( b \), or receive a proof that at least one edge on the way to Bob was broken within \( b \), or receive a streaming payment proportional to time passed beyond \( b \) during which neither was received.

    This mechanism allows for building downstream applications that require provable responses from other participants, such as decentralized storage solutions, decentralized AI agents, and more.
\end{abstract}

\section{Motivation}

It is now inevitable that, in the future, the majority of tasks will be performed by a massive decentralized network of interconnected agents powered by large language models.

As the complexity of the tasks performed by agents increases, so will the average number of dependencies of the agents on other agents. As an example from an adjacent field, Python, NPM, and Rust packages can have dependencies on other packages, and installing a single package, such as a package to compute SHA-256 hashes, often results in the installation of dozens of other packages that it depends on.

Unlike packages, which are generally hosted on a centralized server, and thus the availability of which is relatively reliable, agents will be hosted on a wide variety of hardware, and with varying reliability of implementations. If Alice wants her agent to depend on another agent hosted by Bob, she needs to be certain that the infrastructure on which Bob is hosting their agents is reliable, and that the agent itself does not go down due to software bugs.

If Bob's agent is down, Alice either needs to incur downtime herself; or have her agent's performance degrade due to unavailability of the features provided by Bob's agent; or design her service with potential downtime of Bob's agent, having a fail-over option.

The same problem exists outside of agents, with products and services generally depending on external APIs provided by third parties. Generally, the third party would provide a service-level agreement (SLA) that specifies some guaranteed uptime per unit of time, and if the SLA is not met, the third party generally pays some agreed-upon penalty. We want to generalize this concept to decentralized networks. The problem with such a generalization is that the two entities might not (and generally will not) agree on the duration of the downtime (or whether the downtime happened at all).

A naive solution would be to create a set of validators, such that if Alice claims that Bob's agent or API is down, the validators can verify the claim. With good Sybil resistance, such a system might attest to the availability of services with a high degree of correctness.

The issue with such a naive solution is that it only works if Bob is offline for everyone, including Alice. With increased complexity of services, Bob might be generally online, but could choose to not respond specifically to Alice. For example, Bob might be providing a service that is expensive to compute in certain circumstances and chooses not to respond to Alice because the particular request is too expensive to complete.

Instead, we want to design a system in which for each request Alice issues to Bob, Bob either responds to the request, or a proof is generated that Bob failed to respond, which then can be used to penalize Bob. The particular system we propose does not concern itself with the nature of the requests and responses, nor does it impose any large penalties itself. All it guarantees is that for a particular request signed by Alice, Bob either provides a signed response or can generate a proof – which can be verified on a blockchain – attesting to Bob's failure to respond. Then underlying protocols can introduce penalties either for Bob's message deviating from the expected format or for Bob's failure to respond.

\section{Overview}

The network consists of nodes, and some of the nodes are connected by edges. The entire topology of the network is persisted in a smart contract on NEAR. For each edge in the network, the two nodes have a state channel open (seel \cite{lightning-network} for a good overview of payment channels, and \cite{10.1145/3243734.3243856} for generalized state channels), such that they can exchange small amounts of NEAR, and certain information, without touching the blockchain. Each node has some amount of stake. When the network becomes disconnected, we consider the part that has the smaller cumulative stake to be disconnected from the part with the larger cumulative stake. In such a case, every node in the partition with the smaller stake loses some percentage of their stake. Generally, severing any edge in the network results in a penalty paid by both nodes incident to it. In the event of network partitioning, the nodes in the smaller partition reimburse the nodes in the larger partition for all such penalties incurred.

Each edge \( e \) is a tuple \( \left( L_e, C_e, B_e \right) \). \( L_e \) is the promised latency that is chosen by the two nodes that established the connection. There are no constraints for the latency declared, but there are incentives to declare it close to what the two nodes can actually deliver. \( C_e \) and \( B_e \) define the cost of communicating a message via the edge, with \( C_e \) being the base cost per message, and \( B_e \) being the cost per byte sent.

The main interface the network exposes is an ability for one participant to send a request to another participant, and receive a response. The participants choose the path through the network that provides an acceptable cumulative latency at an acceptable cost, and send the request to the first node on the path.

The entire path does not need to be known to each node on the path. More specifically, each node only needs to know the next node the message needs to be routed to. This functionality is largely orthogonal to the goals of proof of response. We refer the reader to \cite{tor-design} for further details.

The message \( m \) is a tuple \( \left( A_m, D_m, M_m, P_m \right) \), where \( A_m \) is the author of the message, \( D_m \) is the recipient, \( M_m \) is the contents of the message, \( P_m \) is the path along the graph. Let \( L_m \) be the sum of promised latencies via \( P_m \), and \( C_m \) be the sum of the costs to send \( M_m \) via \( P_m \). Let \( B_m \) be the first node on the path \( P_m \). \( A_m \) sends the message to \( B_m \), alongside a payment equal to \( C_m \). Once \( B_m \) acknowledges the message, one of the following things must happen:

\begin{itemize}
    \item The response from \( D_m \) is received by \( A_m \) within \( L_m \);
    \item A proof that at least one edge on \( P_m \) no longer exists in the graph is received by \( A_m \) within \( L_m \);
    \item Neither is received within \( L_m \), but a payment proportional to the extra time that has passed beyond \( L_m \) is paid via the state channel from \( B_m \) to \( A_m \).
\end{itemize}

Since either a response is received, or at least one edge is removed from the graph, \( A_m \) can continue re-sending the messages until either the response is received, or \( D_m \) is disconnected from the network.

Edge operators receive revenue from the payments attached to the messages. As long as the revenue exceeds the stake to open the edge plus the cost of the hardware to maintain the node relaying messages, there is a natural economic incentive to open such edges.

\section{Algorithm}

Consider three nodes \( P \), \( Q \), and \( R \) on some path, such that \( Q \) relayed \( m \) to \( R \), which in turn was relayed to \( Q \) from \( P \).

Consider two outcomes for \( Q \):

\textbf{\( R \) follows the protocol}. In this case, either \( R \) communicated to \( Q \) within the allocated latency the response or the proof of an edge removed from the graph, or \( R \) hasn't communicated either, but sent a sufficient amount of NEAR to pay for the delay. In the former case, \( Q \) is guaranteed to relay the response or the proof of edge removal back to \( P \) in time, assuming the promised latencies between \( R \) and \( Q \), and \( Q \) and \( P \), do not exceed actual. In the latter case \( Q \) relays the entire payment from \( R \) to \( P \). For similar reasons, that fee will be sufficient to cover the delay as seen by \( P \).

\textbf{\( R \) doesn't follow the protocol}. If \( R \) stops following the protocol, \( Q \) submits a transaction on-chain to break the edge between them and \( R \), and sends the proof that such a transaction is sent back to \( P \). A protocol can be designed in such a way that the promise can be sent long before the transaction actually gets finalized on-chain. Note that since an edge removal is much more expensive than the late fees for the message, \( Q \) can choose to wait for some time for \( R \) to respond, while paying the fees to \( P \) out of pocket.

In either case, by induction for as long as \( Q \) follows the protocol, \( P \) will be able to follow the protocol as well, and if \( Q \) stops following the protocol, \( P \) will be able to follow the protocol by severing the edge with \( Q \) and sending the proof of the edge removal upstream, possibly with a delay fee.

\subsection{Payment for severed edges}

For a particular pair of nodes \( P \) and \( Q \), if one of them deviates from the protocol and the other is forced to break the edge, it's generally impossible to tell which node was faulty. Thus, when the edge is broken, by default both nodes will pay a small penalty for the edge removal.

However, if \( P \) has reason to believe that \( Q \) is offline, the following protocol can be implemented to ensure that \( Q \) pays the whole penalty, while \( P \) pays nothing:

\begin{enumerate}
    \item \( Q \) starts deviating from the protocol (refuses to provide response or payment for one or more messages in flight between \( P \) and \( Q \));
    \item \( P \) immediately sends a transaction on-chain to sever the edge, which initially charges both \( P \) and \( Q \) for half of the penalty; \( P \) also sends the proof of edge removal upstream to the previous node on the path;
    \item \( P \) sends a request to \( Q \) to provide the response to the message, or payment for the delay, using Proof of Response;
    \item If \( Q \) is offline, \( P \) will eventually be able to remove a sufficient number of edges from the network to isolate \( Q \), by repeatedly requesting the response. When a node is isolated, the network orchestrating contract on NEAR charges them to reimburse all the fees paid by other nodes for severing edges with them.
\end{enumerate}

\subsection{Bandwidth}

A situation can occur when a particular edge is sufficiently popular that it cannot relay all the messages that choose to be sent through it. Let's say the edge connects nodes \( P \) and \( Q \), and a message arrives to \( P \) to be relayed to \( Q \), but the network between them is at capacity. \( P \) can choose one of the two things:

\begin{itemize}
    \item Queue the message, and pay the fee for the delay;
    \item Break the edge, and reestablish it with higher costs. In this case \( P \) would be able to respond immediately with the proof of an edge removal.
\end{itemize}

In the latter case, the penalty for broken edge must be paid. In other words, the system encourages the nodes to set the cost of relaying the messages to be sufficiently large so that the edge does not get overloaded.

\section{Usage}

Proof of Response itself does not impose any requirements on the contents of the requests and responses nor on the latencies and bandwidths declared.

It is the responsibility of the protocols that are built on top of Proof of Response both to ensure the validity of messages exchanged and to have appropriate requirements for promised latencies and bandwidths.

Participants in Proof of Response only need to provide a small stake that covers the micro-payments necessary for the state channels. Consequently, the protocols that build on top of Proof of Response can require the participants to deposit larger stakes, without making the participants have to stake large amounts twice.

When Alice requests something from Bob via a protocol $B$ built on top of Proof of Response, if the response $r$ is delivered, it is signed by Bob. If Bob deviates from the protocol $B$ in their response, Alice can provide $r$ to the smart contract governing $B$, and such a smart contract will have sufficient information to slash Bob for deviating.

Similarly, if Alice requests something from Bob and Bob does not respond, Alice can repeatedly send the request to Bob until Bob is isolated from the network, and then Alice can provide a proof that Bob is not in the connected component with the largest stake to the smart contract governing $B$, which in turn can slash Bob.

\subsection{Minimum Bandwidth and Maximum Latency}\label{subsection-min-bandwidth-max-latency}
Bob can attempt to bypass this requirement by creating a long chain of edges with very high latency or very low bandwidth, such that technically they are part of the graph, but in practice are not usable. As mentioned above, Proof of Response itself does not impose any constraints on declared latencies or bandwidth. It is the responsibility of $B$ to define what \textit{being available} means. In particular, $B$ can and should define a minimum bandwidth and a maximum latency for the participants in their protocol.

Imposing the minimum bandwidth and maximum latency requires one careful consideration. It could be that Bob is well-connected to the graph, and it is instead Alice who creates a long chain of high-latency edges to the graph, to create a situation in which she and Bob do not have a reasonable connection. For the smart contract governing $B$ to distinguish between such situations, it should not consider the path connecting Bob to Alice, but instead the path connecting Bob to the stake-weighted center of the network. We define the stake-weighted center of the graph in the following way:

$$ \mathrm{argmin}_{v \in V}{\sum_{u \in V, u \ne v}{Distance(u, v) \times Stake(v)}} $$

Where \( V \) is the set of all the nodes, and \( Distance(u, v) \) is the minimum distance, measured in the number of edges traversed, between \( u \) and \( v \).

In other words, the stake-weighted center of the network is such a node, that the average distance to all other nodes, weighted by their stake, is minimized.

Now suppose a participant Eve wants to make a long chain of nodes: $$ v_0, v_1, v_2, v_3, ..., v_n $$ where \( v_0 \) is the first and only node in the chain not controlled by Eve, such that the stake-weighted center of the network ends up landing on one of their nodes (i.e. some node \( v_k, k > 0 \)). Consider the value of the expression under \( \mathrm{argmin} \) in the formula above. Further consider the nodes \( v_k \) and \( v_{k-1} \). Note that for each node \( v_z, z > k \) the value of the expression for \( v_{k-1} \) will increase by the corresponding associated stake, since the distance to such nodes has increased by one, while for every other node in the network the value will decrease by its associated stake, since the distance to such nodes has decreased by one. Thus, for as long as Eve does not control more than half of the stake of the network, for any positive \( k \) the value of the expression for \( v_{k-1} \) will be smaller than for \( v_k \), thus no positive \( k \) \( v_k \) could be the stake-weighted center of the network.

\section{Example Scenarios}

Consider the following topology:

\begin{verbatim}
Alex <- 100ms -> Alice <- 200ms -> Bob <- 100ms -> Carol
                                    ^
Dave <- 100ms --> Eve <-- 100ms ----|
\end{verbatim}

\subsection{Happy case}

\begin{itemize}
    \item Alex requests a packet from Carol. It's 400ms one-way, so he expects the response in 800ms.
    \item Alice receives it 100ms later and forwards it to Bob with 600ms timeout.
    \item Bob receives it 200ms later and forwards it to Carol with 200ms timeout.
    \item Within 200ms, Bob receives the response and sends it back to Alice.
    \item Alice receives it 200ms later. From her perspective, exactly 600ms have passed since she forwarded it.
    \item She sends it to Alex. He receives it 100ms later, exactly 800ms after initiating the request.
\end{itemize}

\subsection{Eve is offline, Bob breaks the edge right away}

\begin{itemize}
    \item Alex requests a packet from Dave. It's 500ms one way, so he expects the response in 1000ms.
    \item Alice receives it 100ms later and forwards it to Bob with 800ms timeout.
    \item Bob receives it 200ms later and forwards it to Eve with 400ms timeout.
    \item 400ms later, Bob doesn't receive the response.
    \item Bob is configured to communicate late payments with Alice every second.
    \item ------ \textit{The above will be common in all remaining examples} --------
    \item Bob doesn't see much value in the edge with Eve. At the end of the 400ms timeout, he immediately initiates the transaction to break the edge with Eve, and sends the proof of the initiation to Alice.
    \item Alice receives it 100ms later and sends it to Alex. Alex has the proof that his request resulted in an edge being removed, and can choose to send the request again, via a different path.
\end{itemize}

\subsection{Eve is offline, Bob waits a bit and pays for it}

\begin{itemize}
    \item Alex requests a packet from Dave. It's 500ms one way, so he expects the response in 1000ms.
    \item Alice receives it 100ms later and forwards it to Bob with 800ms timeout.
    \item Bob receives it 200ms later and forwards it to Eve with 400ms timeout.
    \item 400ms later, Bob doesn't receive the response.
    \item Bob is configured to communicate late payments to Alice every second.
    \item ------ \textit{End of common part} --------
    \item Bob doesn't see much value in the edge with Eve, but decides to wait until their next sync with Alice, which is due in \textapprox 1 second. \textapprox 1 second later, Bob pays Alice out of pocket the delay fee for this \textapprox 1s delay, and initiates the transaction to break the edge with Eve. The message to Alice comprises the delay payment for \textapprox 1s, and the proof of the transaction initiated (so that Alice doesn't expect any more late payments).
    \item Alice is configured to talk to Alex and settle late payments every 0.5 seconds. 500ms after receiving the request from Alex, she is still within 800ms of the timeout. 500ms later, she is 200ms past the timeout. Since she's configured to talk to Bob every second, but to Alex every 0.5 seconds, she will have to communicate a late payment to Alex before she will learn whether or not Bob is about to pay her. This is a risk she needs to be willing to take (otherwise she should not have configured the two edges this way). She pays Alex for the delay, and patiently waits for Bob to send the payment.
\end{itemize}

Here's how this scenario can look. Say that Alice is expected to talk to Alex at 500ms and at 1000ms from the moment she received the request (i.e. their last communication happened to be around the same time as the request), and that she's scheduled to receive the update from Bob at 600ms and 1600ms (i.e. their last communication was \textapprox 400ms ago).

\begin{verbatim}
Time         Alex          Alice           Bob
0          Send request
100                      Forward request
300                                      Forward request
500         <- sync, no late payment ->
600                         <- sync, no late payment ->
700                                      Response due
900                      Response due
1000          <- sync, Alice pays ->
1500          <- sync, Alice pays ->
1600                        <- sync, Bob pays + edge ->
2000        <- sync, Alice pays + edge ->
\end{verbatim}

In the above diagram ``+ edge'' means the inclusion of the proof of broken edge; ``Send request'' and ``Forward request'' are specific to a particular person, and ``<- sync ->'' are between two people.

At 500ms Alice is still within the timeout, so no late payment is expected to be paid to Alex.

Similarly, at 600ms Bob is within the timeout, so no late payment is expected.

At 1000ms, Alice needs to sync with Alex, and she is due for the late payment, but she won't yet know if Bob will pay his late payment – she will only learn by 1600ms. She pays Alex out of pocket.

At 1500ms Alice still doesn't know if Bob will pay her, she has no response from him, and pays Alex out of pocket again.

At 1600ms she receives the payment from Bob, and recoups her losses. She also receives the proof of broken edge, which she forwards to Alex with the remaining late payment. Since everyone now knows that the edge is broken, no more late payments are expected.

\subsection{Eve is offline, Bob waits, but doesn't pay the late fee}

\begin{itemize}
    \item Alex requests a packet from Dave, it's 500ms one way, so he expects the response in 1000ms.
    \item Alice receives it 100ms later and forwards it to Bob with 800ms timeout.
    \item Bob receives it 200ms later and forwards it to Eve with 400ms timeout.
    \item 400ms later, Bob doesn't receive the response.
    \item Bob is configured to communicate late payments with Alice every second.
    \item ------ \textit{End of common part} --------
    \item As in the previous example, Bob doesn't see much value in the edge with Eve, but decides to wait until their next sync with Alice, which is due in \textapprox 1 second. But this time around, \textapprox 1 second later Bob sends the message with the broken edge, pretending he sent it 1 second ago, and doesn't pay the late fee during their sync. By this time, following a similar timeline, Alice has already paid a late fee to Alex twice:
\end{itemize}

\begin{verbatim}
Time         Alex          Alice           Bob
0          Send request
100                      Forward request
300                                      Forward request
500         <- sync, no late payment ->
600                         <- sync, no late payment ->
700                                      Response due
900                      Response due
1000          <- sync, Alice pays ->
1500          <- sync, Alice pays ->
1599                       <- Bob sends proof of edge ->
1600                        <- sync, Bob DOESN'T PAY ->
2000        <- sync, Alice pays + edge ->
\end{verbatim}

This example is not different for Alex, but it is different for Alice, since she lost some money. She cannot prove that Bob's message with the broken edge was delayed. The protocol should not dictate the behavior in this case, but a good heuristic could be "If the amount of money I made from the edge minus the amount of money I lost on situations like this in the past hour has been negative, I will break the edge."

\subsection{Eve is offline, Bob waits longer}

\begin{itemize}
    \item Alex requests a packet from Dave, it is 500ms one way, so he expects the response in 1000ms.
    \item Alice receives it 100ms later and forwards it to Bob with 800ms timeout.
    \item Bob receives it 200ms later and forwards it to Eve with 400ms timeout.
    \item 400ms later, Bob doesn't receive the response.
    \item Bob is configured to communicate late payments to Alice every second.
    \item ------ End of common part --------
    \item In this example, Bob values the edge with Eve, and decides to wait longer. He continues paying Alice.
\end{itemize}

\begin{verbatim}
Time         Alex          Alice           Bob
0          Send request
100                      Forward request
300                                      Forward request
500         <- sync, no late payment ->
600                         <- sync, no late payment ->
700                                      Response due
900                      Response due
1000          <- sync, Alice pays ->
1500          <- sync, Alice pays ->
1600                         <- sync, Bob pays ->
2000          <- sync, Alice pays ->
2500          <- sync, Alice pays ->
2600                         <- sync, Bob pays ->
3000          <- sync, Alice pays ->
3500          <- sync, Alice pays ->
3600                         <- sync, Bob pays ->
...
This continues until eventually Bob decides to give up.
...
9600                        <- sync, Bob pays + edge ->
10000       <- sync, Alice pays + edge ->
\end{verbatim}

From Alice's perspective, during each second she pays Alex out of pocket twice, then gets reimbursed by Bob. That continues until Bob eventually sends her the last payment with the proof of the edge being broken, which she forwards to Alex with the last late payment.

\section{Use case example}

Proof of Response in itself only guarantees that \textit{some} message signed by the destination will be received within the latency of the path, or an edge will be severed, but does not provide further guarantees. Here we outline an example of how this system can be used for a decentralized storage solution.

In decentralized storage solutions, such as Sia, it is relatively simple to ensure that the data was \textit{stored}: the two participants can agree ahead of time on the Merkle root of the data, and then the storage node can periodically get a random offset from a random beacon and provide a Merkle proof for the data stored at that offset.

It is harder to prove that the data was \textit{served}. With Proof of Response, the following scheme can be used:

\begin{enumerate}
    \item\label{item-use-case-premise} Alice, who wants her data to be stored and served, and Bob, who wants to store the data, use a smart contract on NEAR, in which they agree on (a) a Merkle proof of data stored, (b) the latency and bandwidth that Bob promised to deliver, and (c) the uptime that Bob promised to provide.
    \item Whenever Alice needs her data, she uses Proof of Response to fetch it from Bob. If Bob is online, the data is sent to Alice. If Bob is offline, an edge will be broken. Alice can repeatedly query Bob, until he gets disconnected from the network. The smart contract allows Alice to report such a situation, and from the moment Bob is disconnected until Bob gets back into the network and pings the smart contract to update his state, he accrues downtime.
    \item If Bob's downtime exceeds the promised value, he gets penalized, and Alice gets paid.
    \item To prevent Bob from attempting to join the network with latency above or bandwidth below the one they promised in section \ref{item-use-case-premise} above, the smart contract implements the protections described in section \ref{subsection-min-bandwidth-max-latency}.
\end{enumerate}

\section{Conclusion}

Proof of Response provides a primitive that enables users of protocols built on top of it to request something from service providers, and either get a response signed by the service provider or a proof that the service provider is not responding, which can later be submitted to a smart contract on a blockchain.

Such a primitive allows one to design autonomous agents and other APIs that can provide verifiable uptime to their users.

\printbibliography

\end{document}